\documentclass[11pt]{article}
\usepackage[latin1]{inputenc}
\usepackage{amsmath, graphicx, natbib, layout, verbatim}
\usepackage{setspace, bigints, rotating, pdflscape}
\usepackage{fancyhdr}
\usepackage{subfig}

\usepackage[hmargin=1cm,vmargin=2.5cm]{geometry}

\title{The Strength of Arcs and Edges in Interaction Networks: \\ Elements of a Model--Based Approach}

\author{Mauricio Sadinle\footnote{Email: msadinle@stat.cmu.edu}\\\\
\textsc{Carnegie Mellon University} }
\date{\today}

\begin{document}
\maketitle

\abstract{When analyzing interaction networks, it is common to interpret the amount of interaction between two nodes as the strength of their relationship.  We argue that this interpretation may not be appropriate, since the interaction between a pair of nodes could potentially be explained only by characteristics of the nodes that compose the pair and, however, not by pair--specific features.  In interaction networks, where edges or arcs are count--valued, the above scenario corresponds to a model of independence for the expected interaction in the network, and consequently we propose the notions of \textit{arc strength}, and \textit{edge strength} to be understood as departures from this model of independence.  We discuss how our notion of arc/edge strength can be used as a guidance to study network structure, and in particular we develop a stochastic blockmodel for directed interaction networks where arc strength is taken as a latent variable.  We illustrate our approach by studying the interaction between the Kolkata users of the myGamma mobile social network.
}

\section{Introduction}

In many scenarios it is possible to count the amount of interaction between individuals or, more generally, between nodes of a network.  This interaction can be either directed or undirected.  Examples of directed interaction networks include communication networks, where we can count the number of text messages, calls, or e--mails sent from individual to individual \citep[e.g.,][]{DiesnerCarley05,Tyleretal05}; and citation networks, where we can record the number of times certain blog links to another blog \citep[e.g.,][]{AdamicGlance05}, or the number of times one author cites another \citep[e.g.,][]{Ding11}.  Undirected interaction networks include collaboration networks, where we can study the number of papers coauthored by two scholars \citep[e.g.,][]{Newman01} or the number of bills cosponsored by legislators \citep{Fowler06}; patient--sharing networks, where we record the number of patients shared by physicians \citep{Barnettetal12}; and tree interaction networks, where we can record, for instance, the number of common fungal species two tree species can host \citep{Mariadassouetal10}.

When studying interaction networks, it seems natural to interpret the amount of interaction between two nodes as the strength of the arc, or of the edge, depending on the relation being directed, or undirected, respectively.  This interpretation is rather common in practice \citep[see, e.g.,][]{Mariadassouetal10,Barnettetal12}, and it appears in textbooks on social network analysis \citep[see, e.g.,][]{WassermanFaust94}.  We argue, however, that the interaction in a network could potentially be explained under a model of independence, where the expected interaction between nodes is modeled using only nodal characteristics, in which case the interpretation of the amount of interaction as the strength of the arc/edge would not be appropriate.  In this article we present a model--based approach to the concepts of \textit{arc strength}, and \textit{edge strength} in directed, and undirected interaction networks, respectively.  We propose to model the interaction between nodes in a way such that departures from a model of independence can be interpreted as arc/edge strength.  The intuition for our approach is that, after controlling the nodal characteristics that account for the interaction in the network (such as gregariousness and attractiveness in directed networks), a larger arc/edge strength should lead to more interaction between nodes.  This approach follows the long--standing tradition of establishing a null model for the network, and then interpreting departures from this null model as network structure.  This tradition goes back at least to the work of \cite{Blau77}, \cite{Rapoport80}, and \cite{StraussFreeman89}, and it has been used more recently by \cite{HeckathornJeffri01}, \cite{Zhengetal06}, and \cite{DiPreteetal11}.

\subsection{Arc/Edge Strength vs. Granovetter's Tie Strength}

The closely related concept of \textit{tie strength} has received a lot of attention in the social sciences literature.  The first definition of the strength of the tie between two individuals was given by \cite{Granovetter73}: ``\textit{the strength of a tie is a (probably linear) combination of the amount of time, the emotional intensity, the intimacy (mutual confiding), and the reciprocal services which characterize the tie}.''   Although this definition has been operationalized via factor analysis \citep{MarsdenCampbell84, Mathewsetal98}, it is not appropriate for the class of networks that we study in this article.  First of all, our approach aims to study arc/edge strength using the observed interaction between nodes, whereas Granovetter's definition typically requires data collected from cross--sectional surveys that aim to measure Granovetter's components of tie strength.  Despite the fact that Granovetter's definition captures different characteristics that lead to consider a tie as strong, it is not applicable to networks where the nodes are not individuals.  For instance, in blog citation networks, patient--sharing networks, or in tree interaction networks, as mentioned above, it is not clear what ``emotional intensity'' or ``intimacy'' would mean.  Furthermore, Granovetter's definition does not control for nodal characteristics that may lead to more interaction (or more time spent) between nodes, such as node's gregariousness or attractiveness.  Finally, Granovetter's definition implies that the tie from node $i$ to $j$ is as strong as the tie from $j$ to $i$.  We believe, however, that it is more natural to consider asymmetric definitions of strength, i.e., we should allow the strength from $i$ to $j$ to be different than from $j$ to $i$, in particular when the interaction between nodes is directed.  Thus, in this article we use the terms \textit{arc strength}, and \textit{edge strength} to avoid confusion with Granovetter's approach.

\subsection{Online Social Networks}\label{ss:OSN}

\begin{figure}[!t]
\centering
  \centerline{\includegraphics[trim = 1cm 0cm 1cm .5cm, width=.6\linewidth]{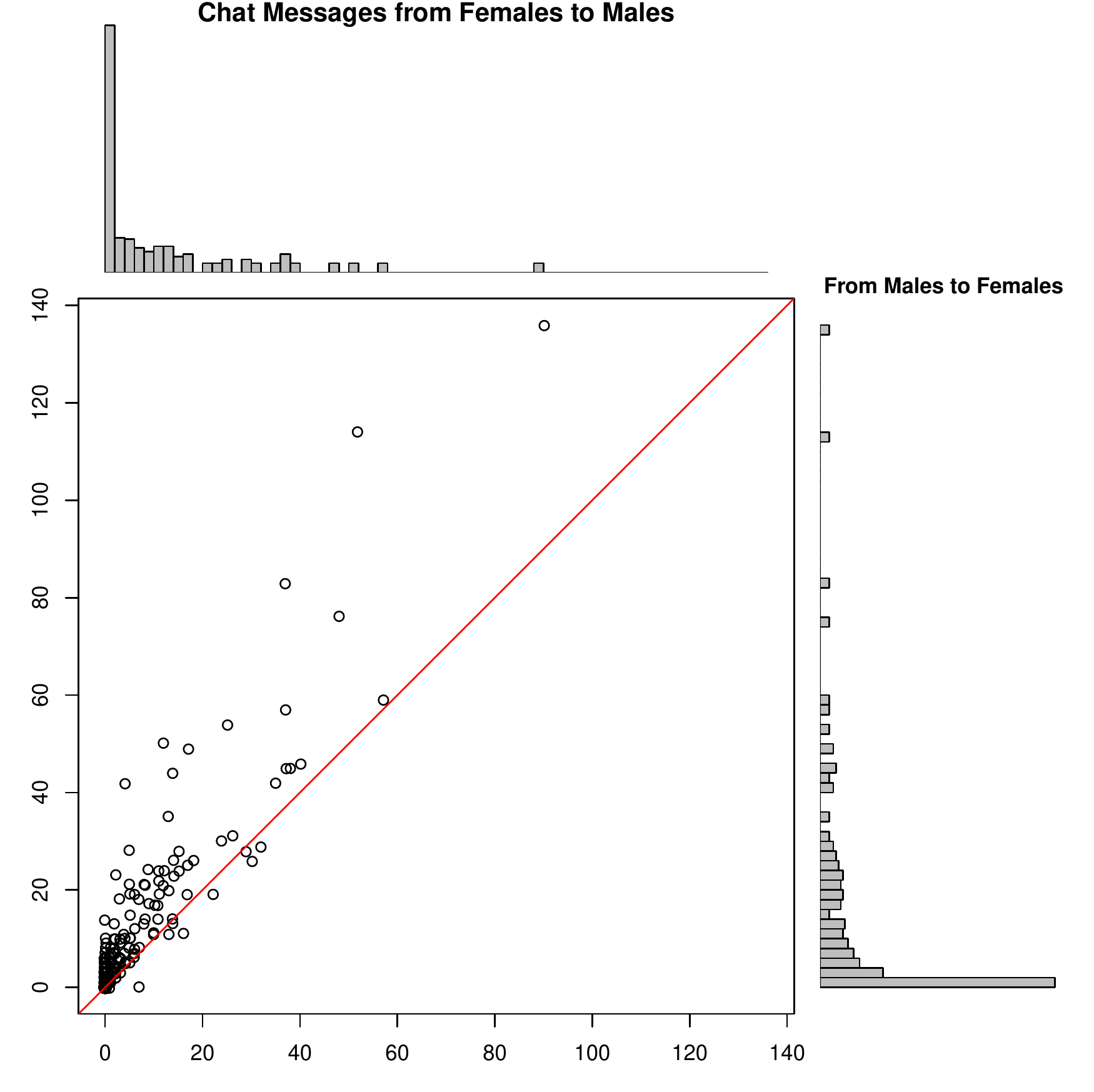}}
  \begin{minipage}[b]{0.85\textwidth}
  \caption{Number of chat messages exchanged between 786 mixed--gender pairs of myGamma users with declared friendships.  These users were located in Kolkata, India, and the chat messages were sent during June 2010.  In the scatterplot, the horizontal axis represents the interaction going from females to males, the vertical axis the interaction from males to females, and the diagonal line represents equality between the number of messages exchanged by the pair of users.  The bars of the histogram on the top represent the frequencies (on the squared root scale) of female--to--male arcs with the number of messages presented in the horizontal axis of the scatterplot.  The histogram on the right has a similar construction for male--to--female arcs.
  }
\label{f:sex_chats}\end{minipage}
\end{figure}

Online social networks, such as \textit{Facebook, Google+}, or \textit{LinkedIn}, offer services focused on facilitating the interaction between users.  We explore the ideas presented in this article using data from the myGamma mobile network.  myGamma (\verb"http://m.mygamma.com/") is a mobile social networking service provided by BuzzCity, a Singapore based company. \cite{BuzzCity07} characterized the users of myGamma as people who access the Internet primarily via mobile phones, living in emerging markets or working in the blue collar sector in wealthier nations.  Users declare friends and foes as directed links, and they interact via chats, messages, blogs, groups, games, etc.  For the purposes of this article, we take users located in the city of Kolkata, India, and we focus on users who were using the networking service prior to May 2010 and who were active during June 2010.  In order to create a count interaction network, we take the number of chat messages sent during June 2010.  In Figure \ref{f:sex_chats}, we present the scatterplot of the interaction between genders.  This plot contains the 786 mixed--gender dyads with declared friendships (we exclude pairs with declared foe links, and one extremely outlying pair with more than 1,000 chat messages exchanged).  The horizontal axis represents the chat messages going from females to males, and the vertical axis the chat messages going from males to females.  The histograms located at each side represent the observed marginal distributions of interaction, and the diagonal line dividing the scatterplot has a slope of one and zero intercept.  Figure \ref{f:sex_chats} shows that only a few pairs of users with declared friendships have large amounts of interaction, whereas most of the interaction among them is null or pretty small.  This distributional characteristic seems to be ubiquitous in online social networks \citep[see, e.g.,][]{GilbertKarahalios09,Xiangetal10}.  Furthermore, the asymmetry between males and females in terms of their amounts of interaction is apparent in Figure \ref{f:sex_chats}, since males tend to send more chat messages to females than females to males.  In Section \ref{s:LASSB} we use our approach to arc strength to construct a model for interaction counts among dyads, and we use it to study the distribution of arc strength among the pairs of Kolkata users with declared friendships in the myGamma network.

\subsection{Overview of the Article}

In Section \ref{s:arcedgestr} we describe in detail our ideal approach to edge and arc strength.  This description contains in Section \ref{ss:nullmodel} the proposal of a null model that can be interpreted as a scenario where no differential values of arc strength are present.  In Section \ref{ss:depindep} we show which are our ideal parameters of arc strength, but we argue that such approach is not feasible, and so in Section \ref{ss:modelaltern} we present a discussion on possible alternatives that conserve the nature of our ideal approach.  In Section \ref{ss:undirected} we briefly present our approach to edge strength for undirected interaction networks, although in the remainder of the article we focus on directed interaction networks.  We introduce in Section \ref{s:LASSB} the \textit{latent arc strength stochastic blockmodels} (LASSB) as a sensible alternative to our ideal approach to arc strength.  In Section \ref{s:appKolkata} we use the LASSB to study the distribution of arc strength in the myGamma Kolkata network presented in Section \ref{ss:OSN}, and in Section \ref{ss:GoF} we present a simulation study to assess the goodness of fit of our model.  Section \ref{s:discussion} contains some conclusions, and a discussion on issues of our approach.  Finally, in Appendix \ref{a:EM} we present an EM algorithm to fit the LASSB via maximum likelihood.

\section{The Strength of Arcs and Edges}\label{s:arcedgestr}

We focus on a network formed by a set of $n$ nodes (a.k.a. vertices) labeled $\{1,\dots,n\}$, and pairs of counts $\{(X_{ij},X_{ji}), i< j\}$ associated to the set of {\em dyads} (pairs of nodes) $\{\{i,j\}, i< j\}$.  The count $X_{ij}$ could be obtained, for instance, as the amount of interaction going from node $i$ to node $j$ during certain period of time, and so we call it \emph{interaction count}.  In undirected interaction networks we call $(i,j)$ the \emph{edge} associated to the dyad $\{i,j\}, i<j$, and $X_{ij}=X_{ji}$ is the value of the edge.  For directed interaction networks, $X_{ij}$ may be different from $X_{ji}$, and each dyad $\{i,j\}, i<j$, has two associated \emph{arcs}: $(i,j)$, and $(j,i)$, which take the values $X_{ij}$, and $X_{ji}$, respectively.

In Sections \ref{ss:nullmodel} and \ref{ss:depindep} we present our guideline approach to arc strength for directed interaction networks, and in Section \ref{ss:undirected} we briefly adapt these ideas to edge strength in undirected networks.

\subsection{A Null Model for Directed Interaction Networks}\label{ss:nullmodel}

In order to construct a model--based approach to arc strength, we need to think of a null model under which we can say that all arcs have the same strength.  Assume the interaction counts $\{(X_{ij},X_{ji}), i< j\}$ follow a distribution $G$, such that $E(X_{ij})=\theta\alpha_{i}\beta_{j}$, for all $i\neq j$, where $\theta$, $\alpha_i$, and $\beta_j$ are positive numbers.  This represents a model of independence for the expected interaction counts, i.e., the expected amounts of interaction are explained only by nodal characteristics.  In this context we call the parameters $\alpha_{i}$ the \textit{gregariousness} of node $i$, $\beta_{j}$ the \textit{attractiveness} of node $j$, and $\theta$ the \textit{density} of interaction, although their specific interpretation is subject to constraints imposed to ensure identifiability.  We believe that a model of independence for the mean interaction is a fundamental component of a scenario that can be interpreted as all arcs having the same strength.

Our null model, however, still needs to completely specify the distribution $G$.  In order to select this distribution, let us think of a process where we can say that there are no differential values of arc strength involved.  Suppose we observe the interaction from node $i$ to node $j$ in the time interval $[0,1]$, so that at time 0 there is no observed interaction.  Assume the amounts of interaction for two non--overlapping time intervals are independent.  Furthermore, assume in any time interval there is a non--zero probability of interaction, and that different single interaction events cannot happen at the same time.  Finally, suppose the interaction counts for two equal sized time intervals are identically distributed.  The reader may realize that we just described a Poisson process \citep[see, e.g.,][]{Parzen62}.  Assuming that we have one independent process for each arc, we obtain a scenario where the nodes interact independently with each other over time.  The interaction count $X_{ij}$ at the end of the period follows a Poisson distribution with certain mean $m_{ij}$.  If these means can be expressed as $m_{ij}=\theta\alpha_{i}\beta_{j}$ for all $i\neq j$, then we say that there are no differential values of arc strength governing the interaction in the network.

The Poisson model with independence for the mean has been taken as a null model in different studies, where departures from it are interpreted as social structure \citep{Zhengetal06,DiPreteetal11}.  This is also one of the simplest models that we can think of when modeling count data, although we can expect to find other more appealing null models of independence for interaction networks.  We expect that the ideas presented here can be adapted easily to those scenarios, but for the remainder of this article we focus on the Poisson model with independence for the mean as the scenario having no differential values of arc strength.

\subsection{Arc Strength as a Departure from Independence}\label{ss:depindep}

From the previous section we have that, if the interaction data $\{(X_{ij},X_{ji}), i< j\}$ are generated independently as \begin{equation}\label{eq:Poisindep}
X_{ij} \overset{ind}{\sim} \text{Poisson}(\theta\alpha_{i}\beta_{j}) \text{ for } i\neq j,
\end{equation}
then we say that all arcs have the same strength in the network.  Notice that although we refer to model \eqref{eq:Poisindep} as a model of independence, it actually corresponds to a quasi--independence model \citep[see][]{Bishopetal75}, since the counts $X_{ii}$, $i=1,\dots,n$, are not defined.  Model \eqref{eq:Poisindep} is not identifiable unless we impose a set of constraints on the sets of parameters $\{\alpha_{1}, \dots, \alpha_{n}\}$, and $\{\beta_{1},\dots,\beta_{n}\}$.  We could fix, for instance, $\alpha_{1}=\beta_{1}=1$, or require $\prod_i \alpha_{i}=\prod_j\beta_{j}=1$, although the set of constraints is arbitrary, and the interpretation of the parameters changes accordingly.  Notice that the model in equation \eqref{eq:Poisindep} has $2n-1$ free parameters after constraints have been imposed, and the amount of valued arcs is $n(n-1)$, which leads to $n^2-3n+1$ degrees of freedom that, in principle, could be used to capture departures from this model.  Let $\lambda_{ij}$ measure the multiplicative departure of $m_{ij}$ from the mean in the independence model of equation \eqref{eq:Poisindep}, i.e., $\lambda_{ij}$ is a parameter included specifically for the arc $(i,j)$, and it can be written as $\lambda_{ij}=m_{ij}/\theta\alpha_{i}\beta_{j}$.  Thus, we say that under the model
\begin{equation}\label{eq:Poisarc}
X_{ij} \overset{ind}{\sim} \text{Poisson}(\theta\alpha_{i}\beta_{j}\lambda_{ij}) \text{ for } i\neq j,
\end{equation}
$\lambda_{ij}$ can be interpreted as the strength of the arc $(i,j)$.  One parameter $\lambda_{ij}$ for each arc represents our ideal measure of arc strength.  This approach, however, is unfeasible.  In order for the $n(n-1)$ parameters $\lambda_{ij}$ to be included jointly with $\{\alpha_i,\beta_i; i=1,\dots,n\}$, we require constraints like $\lambda_{i1}=\lambda_{1j}=1$, or $\prod_i\lambda_{ij}=\prod_j\lambda_{ij}=1$, for all $i$ and $j$, which leads to $(n-1)(n-2)$ free parameters $\lambda_{ij}$.  Consequently, the number of free parameters of model \eqref{eq:Poisarc} is $1+2(n-1)+(n-1)(n-2) = n(n-1)+1$, which exceeds the $n(n-1)$ interaction counts available to us.  We thus conclude that model \eqref{eq:Poisarc} is not suitable for statistical inference in this context.  Model \eqref{eq:Poisarc}, however, represents a guideline for how arc strength should be conceptualized, this is, as a departure from a model of independence.

\subsection{Modeling Alternatives}\label{ss:modelaltern}

The above mentioned difficulties lead us to consider more parsimonious models that conserve the essence of our intuition for arc strength.  The spectrum of alternatives start with modeling parameters as functions of covariates (if available): $\alpha_i$ and $\beta_j$ as functions of nodal covariates, and $\lambda_{ij}$ as a function of arc covariates.  On the other side of the spectrum we have models that take parameters as latent variables: $\alpha_i$ and $\beta_j$'s distributions depend on nodal covariates, and $\lambda_{ij}$'s distribution depends on arc covariates.  The selection of the appropriate model depends on the research purpose, and on the data at hand.  For instance, if the researcher's focus is on exploring the distribution of arc strength, gregariousness, or attractiveness in the network, then modeling these characteristics as latent variables would be the natural way to proceed.  The approach presented in \cite{Xiangetal10} to construct a notion of relationship strength uses a combination of these alternatives, although their input data is a number of dichotomized interaction variables which measure whether specific kinds of interaction are null or not.

An additional motivation for moving towards simplified models is that we may be interested in studying arc strength for only a subset of dyads.  As an example, in online social networks we may want to study arc strength for the subset of dyads with declared binary links (e.g., ``friendships''), in which case the estimation of arc strengths as fixed effects would be even more cumbersome, since the number of interaction counts would be smaller than $n(n-1)$.  Furthermore, in this scenario we may not even be able to fit model \eqref{eq:Poisindep}.  If for instance nodes $i$ and $j$ had declared binary links only among themselves, then only the two interaction counts $X_{ij}$ and $X_{ji}$ would be available from this pair, but model \eqref{eq:Poisindep} would still require the estimation of $\alpha_i$, $\alpha_j$, $\beta_i$, and $\beta_j$.  In order to tackle this scenarios, Section \ref{s:LASSB} presents a simple model that falls somewhere in the middle of the spectrum of possibilities presented above, and it is motivated by the study arc strength in online social networks.

\subsection{Edge Strength in Undirected Interaction Networks}\label{ss:undirected}

In the case of undirected networks, if the interaction counts $\{X_{ij}, i< j\}$ are generated according to the model
\begin{equation}\label{eq:Poisindepundirected}
X_{ij} \overset{ind}{\sim} \text{Poisson}(\theta\beta_{i}\beta_{j}), \text{ for } i< j,
\end{equation}
then we say that there are no different values of edge strength governing the interaction in the network.  We can fix $\beta_1=1$, or require $\prod_j\beta_j=1$, to ensure the identifiability of this model.  Notice that in undirected networks we do not obtain parameters associated with gregariousness or attractiveness.  The model in equation \eqref{eq:Poisindepundirected} is the Poisson analog of the so called \textit{beta model} for undirected binary networks \citep[see][]{Rinaldoetal11, Chatterjeeetal11}.  Similarly as for directed networks, our ideal parameter of edge strength is a multiplicative departure of the expected amount of interaction from the mean in equation \eqref{eq:Poisindepundirected}, i.e., $\lambda_{ij}=m_{ij}/\theta\beta_{i}\beta_{j}$.  The inclusion of the $\lambda_{ij}$ parameters jointly with the $\beta_{i}$ parameters would require an additional set of constraints.  For instance, if we fix $\beta_1=1$, it would suffice to set $\lambda_{1i}=1$, for all $i>1$.  The resulting model would involve $1+(n-1)+[n(n-1)/2 - (n-1)] = n(n-1)/2+1$ free parameters, whereas the number of interaction counts is only $n(n-1)/2$.  This inconvenience leads us to find more parsimonious alternatives, as explained in the previous section.  For the remainder of this article we focus, however, on an alternative model to study arc strength in directed interaction networks.

\section{Latent Arc Strength Stochastic Blockmodels}\label{s:LASSB}

We use our ideal approach to arc strength as a guideline, and propose a pair--dependent stochastic blockmodel \citep{Hollandetal83} to study the distribution of arc strength in a network using count interaction data.  This model assumes that the nodes are divided into homogeneous groups, or blocks, in the sense that the nodes are equally gregarious and attractive within block, and the distribution of arc strength depends only on the nodes' memberships to the different groups.  In other words, this model assumes that the nodes' block--memberships determine the distribution of the dyads' interaction.  This approach aims to model parsimoniously interaction networks not only via a block--structure, but also by treating arc strength as a latent variable.  We call this class of models \textit{latent arc strength stochastic blockmodels} (LASSB).  Notice that we would still need a mechanism for specifying or finding the blocks mentioned above.  The parametrization of the LASSB presented below could be incorporated into the general methodology for block discovery presented in \cite{Mariadassouetal10}.  In this article, however, we assume that the groups can be built up from nodal covariates, where, for instance, the blocks could be specified a priori by the researchers according to their exploratory interests.

We propose to model the distribution of arc strength for each arc--block.  In order to choose a sensible parametrization, we take into account what we have learnt from observing interaction in online social networks.  In Figure \ref{f:sex_chats} we explored the distribution of chat messages for a subset of dyads with declared friendship links in the myGamma mobile social network.  We saw that there are only a few user--arcs with large amounts of interaction, whereas most of the interaction is null or pretty small.  We believe this is evidence that only a few arcs are strong, whereas most of them are weak.  Hence, we propose to model $\lambda_{ij}$ using some distribution defined on the non--negative reals, with a monotonically decreasing probability density function, which indicates that the proportion of arcs decreases as their strength increases.  Furthermore, we observed the interaction from user $i$ to user $j$ to be highly correlated with the interaction from $j$ to $i$.  We propose to capture this feature by allowing correlation of the pair of arc strengths associated to a dyad, and so we model $\lambda_{ij}$ and $\lambda_{ji}$ jointly.

\subsection{Model Description}

Let the $n$ nodes of the network be partitioned into $S$ blocks (or node--blocks) denoted $B_s$, $s=1,\dots,S$.  We say the arc $(i,j)$ belongs to the arc--block $B_{rs}$ if $i\in B_r$ and $j\in B_s$.  Similarly, we say the dyad $\{i,j\}$ belongs to the dyad--block $B_{r\wedge s}$ if $i\in B_r$ and $j\in B_s$, or if  $j\in B_r$ and $i\in B_s$.  Notice that if $(i,j)\in B_{rs}$, then $\{i,j\}\in B_{r\wedge s}$, and $(j,i)\in B_{sr}$.  In particular, notice that if $(i,j)\in B_{rr}$, then $(j,i)\in B_{rr}$.

The approach presented in this article models $\{(\lambda_{ij}, \lambda_{ji}), i<j\}$ indirectly, adapting the ideas of \cite{Nelson85}.  Let the \textit{dyad strength} be
$\lambda_{i \wedge j} = \lambda_{ij}+ \lambda_{ji}$, and let the \textit{arc share} be $\rho_{ij} = \lambda_{ij}/\lambda_{i \wedge j}$.  From this formulation, the closer $\rho_{ij}$ to 0.5, the larger the reciprocity in the relationship between nodes $i$ and $j$.  Note that we can obtain back $\lambda_{ij} = \rho_{ij}\lambda_{i \wedge j}$, and $\lambda_{ji} = (1-\rho_{ij})\lambda_{i \wedge j}$.
By using this transformation in our block--modeling approach, the distribution of $\lambda_{i \wedge j}$ depends only on the dyad $\{i,j\}$'s membership to the different dyad--blocks, and the distribution of $\rho_{ij}$ depends only on the membership of the arc $(i,j)$ to the different arc--blocks.  Notice that the distribution of $\rho_{ij}$ trivially determines the distribution of $\rho_{ji}=1-\rho_{ij}$.  Consequently, for modeling $\{(\lambda_{ij}, \lambda_{ji}), i<j\}$ we need to specify the distribution of $\lambda_{i \wedge j}$ for each of the $S(S+1)/2$ dyad--blocks, and the distribution of $\rho_{ij}$ for each of the $S(S+1)/2$ arc--blocks $B_{rs}$ with $r\leq s$.

Modeling $\lambda_{i \wedge j}$ as a gamma random variable allows the marginal distribution of $\lambda_{ij}$ to have the desired characteristics that we mentioned before, since the gamma density function is monotonically decreasing when its shape parameter $\nu_{r\wedge s}$ is lower than or equal to one.  In this article we thus propose a parametrization of the LASSB for the interaction data $X:=\{(X_{ij},X_{ji}), i< j\}$ using a hierarchical structure, as follows:
\begin{eqnarray}\label{eq:LASSB}
\lambda_{i\wedge j}| (i,j)\in B_{rs} &\overset{iid}{\sim}& \text{Gamma}(\mu_{r\wedge s},\nu_{r\wedge s}),\\
\rho_{ij}| (i,j)\in B_{rs} &\overset{iid}{\sim}& \text{Beta}(\pi_{rs},\phi_{r\wedge s}),\nonumber\\
X_{ij}|\lambda_{ij}, (i,j)\in B_{rs} &\overset{ind}{\sim}& \text{Poisson}(\theta\alpha_{r}\beta_{s}\lambda_{ij}),\nonumber\\
X_{ji}|\lambda_{ji}, (i,j)\in B_{rs} &\overset{ind}{\sim}& \text{Poisson}(\theta\alpha_{s}\beta_{r}\lambda_{ji}),\nonumber
\end{eqnarray}
for all $i<j$, and $r\leq s$.  Without loss of generality, we assume that the ordering of the nodes is such that if $i\in B_r$, $j\in B_s$, and $r<s$, then $i<j$.  The gamma and beta parts of the model are parameterized in terms of their means $\mu_{r\wedge s}$, and $\pi_{rs}$, respectively (their densities are presented at the beginning of Appendix \ref{a:EM} for clarification).  Note that if $\rho_{ij}\sim\text{Beta}(\pi_{rs},\phi_{r\wedge s})$, then $\rho_{ji}\sim\text{Beta}(\pi_{sr},\phi_{r\wedge s})$, with $\pi_{sr}=1-\pi_{rs}$.  For dyad--blocks $B_{ss}$ we fix $\pi_{ss}=0.5$, since the ordering of the nodes within the same block is arbitrary.

It is easy to see that if the parameters $\{\theta, \alpha_{r}, \beta_{s}, \mu_{r\wedge s}; r, s=1,\dots,S\}$ are not jointly constrained, then the model in equation \eqref{eq:LASSB} is not identifiable.  Let us then introduce an equivalent representation of the model.  Let $\lambda_{i \wedge j}=\gamma_{i\wedge j}\mu_{r\wedge s}$, for $\{i,j\} \in B_{r\wedge s}$, namely $\gamma_{i\wedge j}$ measures how dyad strength departs from its mean value, and so we call it \textit{relative dyad strength}.  Defining $\gamma_{ij}=\rho_{ij}\gamma_{i\wedge j}$ and $\gamma_{ji}=(1-\rho_{ij})\gamma_{i\wedge j}$, the following parametrization is equivalent to the one presented in equation \eqref{eq:LASSB}:
\begin{eqnarray}\label{eq:LASSB1}
\gamma_{i\wedge j}| (i,j)\in B_{rs} &\overset{iid}{\sim}& \text{Gamma}(1,\nu_{r\wedge s}),\\
\rho_{ij}| (i,j)\in B_{rs} &\overset{iid}{\sim}& \text{Beta}(\pi_{rs},\phi_{r\wedge s}),\nonumber\\
X_{ij}|\gamma_{ij}, (i,j)\in B_{rs} &\overset{ind}{\sim}& \text{Poisson}(\theta\alpha_{r}\beta_{s}\mu_{r\wedge s}\gamma_{ij}),\nonumber\\
X_{ji}|\gamma_{ji}, (i,j)\in B_{rs} &\overset{ind}{\sim}& \text{Poisson}(\theta\alpha_{s}\beta_{r}\mu_{r\wedge s}\gamma_{ji}),\nonumber
\end{eqnarray}
for all $i<j$, $r\leq s$.  The expression $\theta\alpha_{s}\beta_{r}\mu_{r\wedge s}$ represents a quasi--symmetry model \citep[see][]{Bishopetal75,Agresti02} at the block level, since $\mu_{r\wedge s}=\mu_{s\wedge r}$ given that $B_{r\wedge s}\equiv B_{s\wedge r}$.  Hence, we set the constraints $\alpha_1=1, \beta_1=1$, and $\mu_{1\wedge s}=1$ for $s=1,\dots,S$, to avoid non--identifiability of the model.

In model \eqref{eq:LASSB1} there is one parameter $\nu_{r\wedge s}$, and one $\phi_{r\wedge s}$ for each dyad--block, which lead to $2\times S(S+1)/2$ parameters.  Although there are also $S(S+1)/2$ parameters $\mu_{r\wedge s}$, $S$ of them are constrained to be one.  Model \eqref{eq:LASSB1} also has one parameter $\pi_{rs}$ per arc--block, but $S$ of them are fixed to be 0.5, and the constraints $\pi_{rs}+\pi_{sr}=1$, for all $r<s$, have to hold, which lead to $S(S-1)/2$ free $\pi_{rs}$ parameters.  Finally, model \eqref{eq:LASSB1} includes one parameter $\theta$, and $2\times (S-1)$ free node--block parameters $\alpha_r$ and $\beta_s$.  We thus conclude that the LASSB as parameterized in equation \eqref{eq:LASSB1} involves $2S^2+2S-1$ free parameters.  In Appendix \ref{a:EM} we present an EM algorithm for the estimation of the set of parameters $\Phi=\{\theta,\alpha_r,\beta_s,\mu_{r\wedge s},\nu_{r\wedge s}, \pi_{rs}, \phi_{r\wedge s}; r,s=1,\dots,S\}$ via maximum likelihood.

\subsection{Model Interpretation}

The latent arc strength stochastic blockmodel (LASSB), as presented in equation \eqref{eq:LASSB}, indicates a generative process for the observed dyads' interaction, which only depends on the membership of the arcs to the different arc--blocks.  As presented in equation \eqref{eq:LASSB}, we assume that a gamma distribution generates the strength of the dyad $\{i,j\}$, which is shared between the two corresponding arcs according to a beta distribution.  Given the strength of the arc $(i,j)$, $\lambda_{ij}=\rho_{ij}\lambda_{i\wedge j}$, the interaction from $i$ to $j$, $X_{ij}$, is distributed according to a Poisson distribution, with a mean parameter that depends on the arc strength, and also on the gregariousness of node $i$, and the attractiveness of node $j$, which are assumed to be constant for all nodes within the same block.  An interesting feature of the LASSB is that, thanks to its block structure, it does not necessarily require the complete set of $n(n-1)/2$ dyads for it to be fitted \citep[see][]{NowickiSnijders01}.  In particular, it can be fitted to a subset of dyads that are linked at a basic level, such as connected pairs of users in online social networks.

\subsubsection{Distribution of Dyad Strength}

According to the parametrization of the LASSB, we cannot estimate directly the mean dyad strengths $\{\mu_{r\wedge s}, r\leq s\}$ since they have to be constrained jointly with the gregariousness and attractiveness parameters.  However, the shape parameters $\{\nu_{r\wedge s}, r\leq s\}$ capture important information about the distribution of dyad strength.  The parameter $\nu_{r\wedge s}$ controls the shape of the probability density function (PDF) of dyad strength, and also its variance, since $Var(\lambda_{i \wedge j})=\mu_{r\wedge s}^2/\nu_{r\wedge s}$
for $\{i,j\}\in B_{r\wedge s}$.  As $\nu_{r\wedge s}$ goes to infinity, the PDF of dyad strength gets concentrated around its mean, and its variance goes to zero, indicating that all dyads tend to have the same strength.  On the other hand, as $\nu_{r\wedge s}$ goes to zero, the PDF of dyad strength becomes more skewed to the right, and the variance goes to infinity, indicating that the dyads become more heterogeneous in terms of their strengths. For instance, when studying online social networks we would expect $\nu_{r\wedge s}$ to be small for all dyad--blocks, as we expect to have lots of weak dyads and only a few strong ones.

\subsubsection{Distribution of Arc Share}

In the LASSB we assign an arbitrary ordering to the blocks, which allows to estimate the mean arc share $\pi_{rs}$ of the arc--block $B_{rs}$, for $r<s$ ($r\neq s$), since in those cases we can create ordered pairs $(i,j)$ from nodes $i \in B_r$, and $j \in B_s$.  The arc share indicates how symmetric the relationship between a pair of nodes is, and then $\pi_{rs}$ allows to find how symmetric on average the relationships in each dyad--block are.  For dyads where both nodes belong to the same block, it is not possible to assign a meaningful order to the pair, and consequently we fix $\pi_{ss}=0.5$.  In all cases, however, the shape parameter of the beta distribution $\phi_{r\wedge s}$ controls the concentration of the arc share's PDF around its mean.  Furthermore, $\phi_{r\wedge s}$ controls the variance of arc share, since $Var(\rho_{ij})=\pi_{rs}\pi_{sr}/(\phi_{r\wedge s}+1)$, for $(i,j)\in B_{rs}$.  Hence, large values of $\phi_{r\wedge s}$ indicate that most arcs in $B_{rs}$ tend to have the same share in their corresponding dyad strengths, or in other words, most relationships of dyads in $B_{r\wedge s}$ tend to be equally symmetric or equally asymmetric, depending on the value of $\pi_{rs}$. On the other hand, low values of $\phi_{r\wedge s}$ indicate a large spread of arc share, or equivalently, large heterogeneity in terms of how symmetric or asymmetric the dyads' relationships are.

\subsubsection{Association Measures}

Even though the LASSB controls gregariousness and attractiveness per block, the parameters $\{\theta, \alpha_{r}, \beta_{s}, \mu_{r\wedge s}; r,s=1,\dots,S\}$ are not interpretable directly, given that the set of constraints on them is arbitrary.  Nevertheless, these parameters determine some measures that are informative of network structure.  For instance, let $m_{rs}:=\theta\alpha_{r}\beta_{s}\mu_{r\wedge s}\pi_{rs}$ be the marginal expected amount of interaction for arcs in $B_{rs}$.  The ratio $m_{rs}/m_{sr}$ compares the frequency of interaction from block $B_r$ to block $B_s$ with respect to the interaction from $B_s$ to $B_r$, and it can be written as
\begin{equation*}
\frac{m_{rs}}{m_{sr}} = \frac{\alpha_r}{\alpha_s}\frac{\beta_s}{\beta_r}\frac{\pi_{rs}}{(1-\pi_{rs})},
\end{equation*}
where it is clear that this \emph{interaction ratio} is determined by how gregarious nodes in $B_r$ are with respect to nodes in $B_s$, how attractive nodes in $B_s$ are with respect to nodes in $B_r$, and how asymmetric on average the relationships of dyads in $B_{r\wedge s}$ are, which is represented by the odds $\pi_{rs}/(1-\pi_{rs})$.  Notice however, that $\pi_{rs}/(1-\pi_{rs})$  does not depend on different sets of constraints.  Consequently, the ratio
\begin{equation*}
\frac{m_{rs}/m_{sr}}{\pi_{rs}/(1-\pi_{rs})} = \frac{\alpha_r}{\alpha_s}\frac{\beta_s}{\beta_r}
\end{equation*}
is invariant to different constraint sets for the gregariousness and attractiveness parameters, and it can be interpreted as discounting the average asymmetry of the relationships from the interaction ratio, which leads to a measure determined only by the gregariousness and attractiveness of the blocks.  Another interesting measure is the \emph{block--odds ratio}
\begin{equation*}
\frac{m_{rs}/m_{rs'}}{m_{r's}/m_{r's'}} = \frac{\mu_{rs}/\mu_{rs'}}{\mu_{r's}/\mu_{r's'}},
\end{equation*}
which depends only on the mean arc strengths per arc--blocks $\mu_{rs} := \mu_{r\wedge s}\pi_{rs}$.  The above block--odds ratio takes block $B_r$'s odds of interacting to $B_s$ vs. $B_{s'}$, and compares them to block $B_{r'}$'s odds of interacting to $B_s$ vs. $B_{s'}$.  Since this measure does not depend on gregariousness nor attractiveness of the blocks, it allows to discover comparative associations between blocks due only to the strength of the relationships between nodes.

\subsubsection{Recovered Arc Shares and Relative Arc Strengths}

The hierarchical structure of the LASSB also allows us to explore $\lambda^{*}_{ij}:=\gamma_{ij}/\pi_{rs}=\lambda_{ij}/\mu_{rs}$ for $(i,j)\in B_{rs}$, which measures how arc strength departs from its expected value, and so we call it \textit{relative arc strength}.  The variance of $\lambda^{*}_{ij}$ can be written as
\begin{equation}\label{eq:varrelarc}
Var(\lambda^{*}_{ij}) = \frac{1}{\nu_{r\wedge s}}\left[\frac{\pi_{rs}\phi_{r\wedge s} + 1}{\pi_{rs}(\phi_{r\wedge s} + 1)}\right] + \frac{1-\pi_{rs}}{\pi_{rs}(\phi_{r\wedge s} + 1)}, \text{ if } (i,j)\in B_{rs}.
\end{equation}
We saw that in scenarios where most relationships are symmetric, or closely symmetric, $\phi_{r\wedge s}$ is large, and $\pi_{rs}$ is close to 0.5.  In those cases the factor within brackets in equation \eqref{eq:varrelarc} would be basically equal to one, and the second summand in \eqref{eq:varrelarc} would be close to zero, which means that $Var(\lambda^{*}_{ij})$ would be controlled by the variability of relative dyad strength, which is $1/\nu_{r\wedge s}$.
The measure $\lambda^{*}_{ij}$ is interesting since it allows to adjust for the homophily/heterophily captured by the blocks.  For instance, suppose two arcs are equally strong, i.e., $\lambda_{ij}=\lambda_{il}$, and that we only have two blocks in the network. Say $i,j\in B_{1}$, $l \in B_{2}$, and $\mu_{11}\gg \mu_{12}$.  This scenario corresponds to one where homophily explains part of the network structure, since nodes in block $B_1$, on average, have stronger arcs with nodes in the same block than with nodes in block $B_2$.  Consequently, under this scenario the arc $(i,l)$ is relatively stronger than $(i,j)$.  Thus, relative arc strength accounts for these scenarios, and it can be recovered as $\hat \lambda^{*}_{ij} = E_{\hat \Phi}(\rho_{ij}\gamma_{i\wedge j}|X=x)/\hat \pi_{rs}$, for $(i,j)\in B_{rs}$, where $E_{\hat \Phi}(\rho_{ij}\gamma_{i\wedge j}|X=x)$ can be obtained as in equation \eqref{eq:Erhogamma} in Appendix \ref{a:EM}, using the maximum likelihood estimates (MLEs) $\hat \Phi$.  Naturally, we can also explore the recovered arc share $\hat\rho_{ij}=E_{\hat\Phi}(\rho_{ij}|X=x)=(x_{ij} +\hat\phi_{r\wedge s}\hat\pi_{rs})/(x_{ij} + x_{ji} +\hat\phi_{r\wedge s} )$, for $(i,j)\in B_{rs}$, and the recovered relative dyad strength $\hat \gamma_{i\wedge j} = E_{\hat \Phi}(\gamma_{i\wedge j}|X=x)$, which can be computed using $\hat\Phi$, and equation \eqref{eq:Egamma} in Appendix \ref{a:EM}.  Notice that in the above notation $x:=\{(x_{ij},x_{ji}), i< j\}$ represents the observed interaction counts.

\section{Friendships in the myGamma Kolkata Network}\label{s:appKolkata}

In this section we present the fit of the LASSB to the myGamma data described in Section \ref{ss:OSN}.  We take chat messages exchanged during June 2010 between dyads that were connected by at least one directional friendship, and we focus on users in Kolkata, India, taking gender as our a priori blocking criterion.  Our data consists of 786 mixed--gender dyads ($b_{F\wedge M} = 786$), 33 dyads of females ($b_{F\wedge F} = 33$), and 156 dyads of males ($b_{M\wedge M} = 156$), which jointly involve 188 different users.  Among this set of users, 32 are involved in isolated dyads (i.e. there are 16 isolated dyads), and so we would not be able to estimate individual gregariousness and attractiveness parameters for each of them.

We fit the LASSB using the EM algorithm presented in Appendix \ref{a:EM}, and in order to obtain confidence limits for the parameters of the model, and for other interesting functions of the parameters, we follow a parametric bootstrap approach.  We generate 300 interaction networks from the maximum likelihood fitted LASSB, and then we fit the LASSB to each of these bootstrap networks.  Each bootstrap network is obtained by sampling interaction from the fitted model for each of the friendship dyads that we study.  Since all the LASSB's parameters are defined on the non--negative reals, we transform them to the natural logarithm scale, then we compute basic bootstrap confidence limits \citep[see][]{DavisonHinkley97} for the log--parameters, and finally we exponentiate the limits of these intervals.  The same procedure was used to find confidence limits for the functions of the parameters that we study, since they are also defined on the non--negative reals.  All these simulations and computations were performed using the software R \citep{R12}.  We present the results in a series of four tables containing MLEs, and parametric bootstrap confidence limits with a 95\% nominal confidence.  Tables \ref{t:EMresults1} and \ref{t:EMresults2} contain the estimates of the LASSB's parameters, Table \ref{t:funcofparameters} contains the estimates for some functions of the parameters that are indicative of network structure, and Table \ref{t:estmeaninter} contains the estimates of mean interaction per arc--block.  The point estimates for the functions of the parameters are obtained using the invariance property of MLEs.  In Tables \ref{t:EMresults1}, and \ref{t:EMresults2} the values with asterisks are pre--fixed, as explained in Section \ref{s:LASSB}.  We also present in Table \ref{t:EMresults2} the number of dyads in the dyad--blocks $B_{r\wedge s}$, which are denoted $b_{r\wedge s}$.

\begin{table}[!t]
\centering
 \vspace*{6pt}
  \begin{minipage}{0.8\textwidth}
 \def\~{\hphantom{0}}
  \caption{MLEs of $\theta$ and node--block parameters in the LASSB for the Kolkata interaction data.  Numbers preceded by an asterisk are fixed in the model.  Parametric bootstrap 95\% confidence limits appear in parenthesis under the corresponding point estimates.} \label{t:EMresults1}
  \centering
\begin{tabular}{r|lrr}\hline
&&&\\[-8pt]
$\hat\theta$ & Block $s$ & $\hat\alpha_s$ & $\hat\beta_s$ \\
&&&\\[-8pt]\hline
&&&\\[-8pt]
1.42         &Female & *1 & *1 \\
\scriptsize{(0.57 -- 7.91)}&       &    &   \\
    &Male & 3.77         &3.80             \\
    &     & \scriptsize{(0.68 -- 9.39)}  & \scriptsize{(0.69 -- 9.55)}\\\hline
\end{tabular}
\end{minipage}
\vspace*{.7cm}
\centering
 \vspace*{6pt}
  \begin{minipage}{0.8\textwidth}
 \def\~{\hphantom{0}}
  \caption{MLEs of arc--block and dyad--block parameters in the LASSB for the Kolkata interaction data.  Numbers preceded by an asterisk are fixed in the model.  Parametric bootstrap 95\% confidence limits appear in parenthesis under the corresponding point estimates.} \label{t:EMresults2}
  \centering
\begin{tabular}{lrrrrr}\hline
&&&&&\\[-8pt]
Arc--Block $rs$ & $\hat\mu_{r\wedge s}$ &$\hat\nu_{r\wedge s}$ & $\hat\pi_{rs}$     & $\hat\phi_{r\wedge s}$ & $b_{r\wedge s}$\\
&&&&&\\[-8pt]\hline
&&&&&\\[-8pt]
Female -- Male  & *1  &       0.10   &    0.27       & 12.62                & 786\\
                &     &\scriptsize{(0.09 -- 0.12)}& \scriptsize{(0.24 -- 0.29)}& \scriptsize{(8.87 -- 17.84)}       & \\
&&&&&\\[-6pt]
Female -- Female& *1  &0.07          &   *0.50       & 56,940.19            & 33\\
                &     &\scriptsize{(0.02 -- 0.15)}&               &\scriptsize{(30,371.28 -- 1,417,551,821)}   & \\
&&&&&\\[-6pt]
Male -- Male   & 0.28         &      0.05    &   *0.50        & 71.32            & 156\\
               &\scriptsize{(0.08 -- 2.13)}&\scriptsize{(0.03 -- 0.07)}&                & \scriptsize{(0.08 -- 239.79)}   & \\\hline
\end{tabular}
\end{minipage}
\end{table}
\begin{table}[t]
\centering
 \vspace*{6pt}
  \begin{minipage}{0.8\textwidth}
 \def\~{\hphantom{0}}
  \caption{Some measures of network structure for the Kolkata interaction data.  We present their MLEs along with parametric bootstrap 95\% confidence limits (CL).} \label{t:funcofparameters}
  \centering
\begin{tabular}{lrc}\hline
&&\\[-8pt]
Measure & MLE & CL \\
&&\\[-8pt]\hline
&&\\[-8pt]
$\nu_{M\wedge M}/\nu_{F\wedge M}$ & 0.45 & 0.28 -- 0.69 \\
&&\\[-8pt]
$\nu_{F\wedge F}/\nu_{F\wedge M}$ & 0.66 & 0.20 -- 1.54 \\
&&\\[-8pt]
$\nu_{F\wedge F}/\nu_{M\wedge M}$ & 1.45 & 0.45 -- 3.71 \\
&&\\[-8pt]
$\phi_{F\wedge F}/\phi_{M\wedge M}$ & 798.38 & 278.49 -- 1,050,115,823 \\
&&\\[-8pt]
$\frac{ m_{MF}/ m_{MM}}{ m_{FF}/ m_{FM}}$ & 2.82 & 0.37 -- 9.68 \\
&&\\[-8pt]
$m_{MF}/m_{FM}$ & 2.69 & 2.38 -- 3.07 \\
&&\\[-8pt]
$\frac{m_{MF}/m_{FM}}{\pi_{MF}/\pi_{FM}}$ & 0.99 & 0.98 -- 0.99 \\
&&\\[-8pt]\hline
\end{tabular}
\end{minipage}
\vspace*{.7cm}
\centering
 \vspace*{6pt}
  \begin{minipage}{0.8\textwidth}
 \def\~{\hphantom{0}}
  \caption{Estimated mean interaction per arc--blocks $m_{rs}:=\theta\alpha_{r}\beta_{s}\mu_{r\wedge s}\pi_{rs}$.  Parametric bootstrap 95\% confidence limits appear in parenthesis under the corresponding MLEs.} \label{t:estmeaninter}
  \centering
\begin{tabular}{lrr}\hline
&&\\[-8pt]
& \multicolumn{2}{c}{To}\\
   \cline{2-3}\\ [-8pt]
From & Female & Male \\
&&\\[-8pt]\hline
&&\\[-8pt]
Female &    0.71               &  1.46 \\
       &\scriptsize{(0.28 -- 3.96)}  & \scriptsize{(1.16 -- 1.91)}\\
Male   &   3.93                &  2.85 \\
       &\scriptsize{(3.07 -- 4.97)}  &  \scriptsize{(1.60 -- 6.23)}\\\hline
\end{tabular}
\end{minipage}
\end{table}

The estimates $\hat \nu_{r\wedge s}$ in Table \ref{t:EMresults2} indicate that dyads involving both genders have less variability in terms of their relative dyad strength, compared to dyads involving the same gender.  For instance, the estimated variability of relative dyad strength for dyads involving only males is estimated as $1/\hat \nu_{M\wedge M}\approx 20$, which doubles the corresponding variability for dyads involving both genders.  The fact that the upper confidence limit for $\nu_{M\wedge M}/\nu_{F\wedge M}$ in Table \ref{t:funcofparameters} is bounded below one allows us to say with 95\% confidence that the variability of relative dyad strength for male--male dyads is larger than the corresponding variability for dyads involving both genders.  The same conclusion, however, can not be stated when comparing the variability of relative dyad strength for female--female dyads with the other two groups of dyads, since the confidence limits for $\nu_{F\wedge F}/\nu_{F\wedge M}$, and for $\nu_{F\wedge F}/\nu_{M\wedge M}$ include one.  The lecture of the estimates $\{\hat \nu_{r\wedge s}; r,s\in \{F,M\}\}$ also tells us that the proportion of relatively weak dyads involving only males tends to be larger, compared to mixed--gender dyads, since the proportion of $\gamma_{i \wedge j}$ being close to zero is larger for, say, $\hat \nu_{M\wedge M}=0.05$, than for $\hat \nu_{F\wedge M}=0.10$.

\begin{figure}[!t]
\centering
  \centerline{\includegraphics[trim = 1cm .1cm 1cm 1cm, width=.85\linewidth]{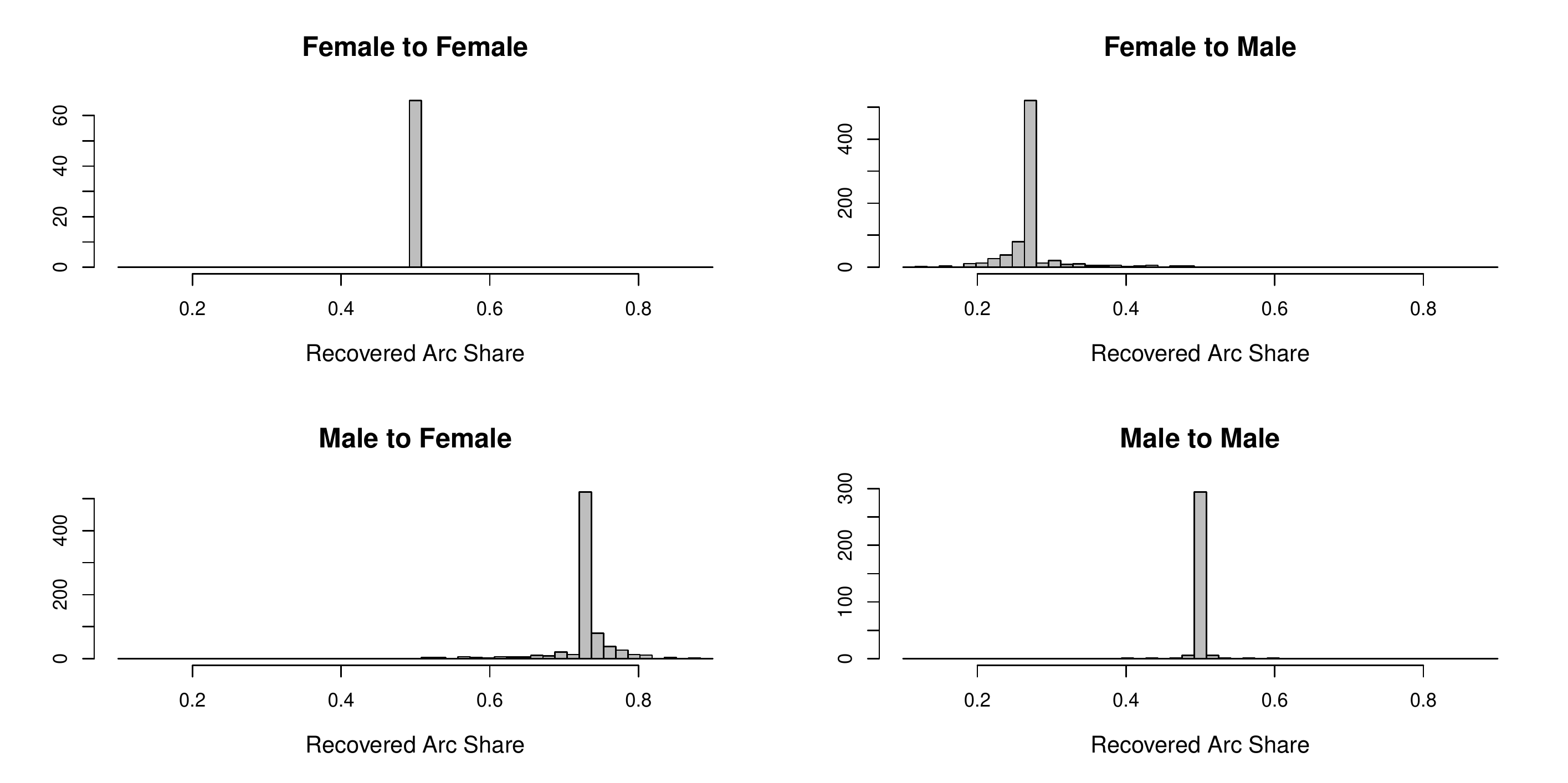}}
  \begin{minipage}[b]{0.8\textwidth}
  \caption{Recovered arc share for the four arc--blocks.
  }\label{f:recovshare}
\end{minipage}
\centering
  \centerline{\includegraphics[trim = 1cm 0cm 1cm 0cm, width=.85\linewidth]{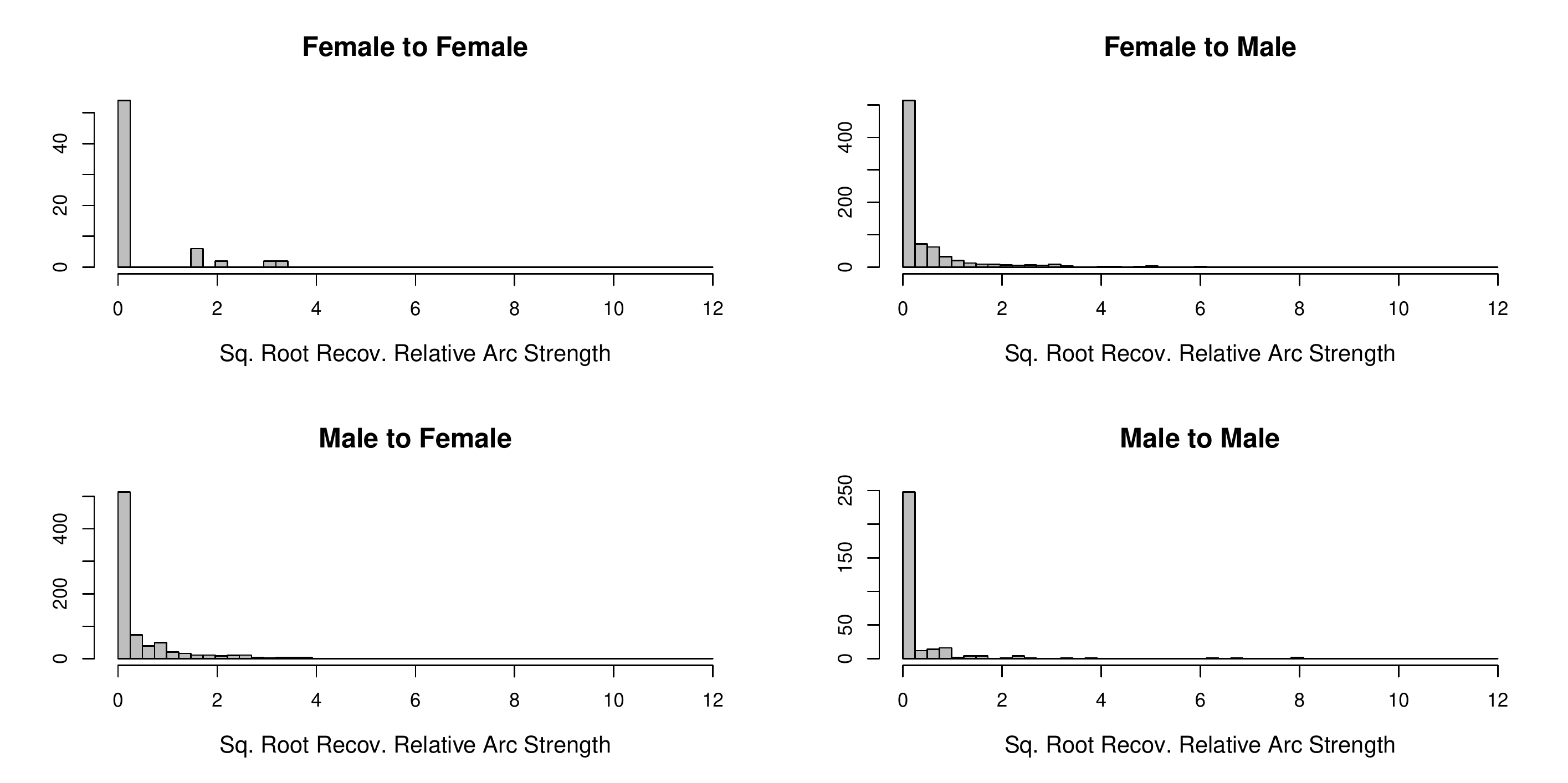}}
  \begin{minipage}[b]{0.8\textwidth}
  \caption{Recovered relative arc strength (on the squared root scale) for the four arc--blocks.
  }\label{f:recovrelarcstr}
\end{minipage}
\end{figure}

We can also see that the relationships between males and females are asymmetric on average, since the estimated mean arc share for female--to--male arcs, $\hat\pi_{FM}$, is $0.27$,  which indicates that out of the dyad strength for mixed--gender dyads, only 27\% on average corresponds to the strength of the female--to--male arc.  The 95\% confidence limits associated to $\pi_{FM}$ are 0.24 -- 0.29 (Table \ref{t:EMresults2}), with which we can reject the hypothesis of average symmetry in the relationships between males and females ($\pi_{FM} = 0.5$).  We can also see that the relationships between females tend to be highly symmetric, since $\hat\phi_{F\wedge F}=56,940.19$ indicates that the arc shares are highly concentrated around $0.5$.  Notice that although the confidence interval for $\phi_{F\wedge F}$ is very wide, the values that it contains support the same claim.   Compared to relationships between females, relationships between males seem to have a wider range of variability in terms of their asymmetries, since $\hat\phi_{M\wedge M}=71.32$ indicates that the arc shares are more spread out from $0.5$.  This conclusion can also be stated from the confidence interval for $\phi_{F\wedge F}/\phi_{M\wedge M}$ in Table \ref{t:funcofparameters}, which is located far above one.  These variabilities are also observed in the recovered $\hat\rho_{ij}$, which are presented in Figure \ref{f:recovshare}.

In equation \eqref{eq:varrelarc} we saw how the variance of relative arc strength behaves as a function of $\nu_{r\wedge s}$, $\phi_{r\wedge s}$, and $\pi_{rs}$.  According to the obtained estimates, we can see that for all blocks, the variability of relative dyad strength dominates $Var(\lambda^{*}_{ij})$, since the parameters $\hat \nu_{r\wedge s}$ are close to zero, and the $\hat \phi_{r\wedge s}$ are large in general.  This variabilities are also reflected in the histograms of $\hat \lambda^{*}_{ij}$, which are presented in Figure \ref{f:recovrelarcstr}.

From Table \ref{t:funcofparameters}, males' odds of sending chat messages to a female vs. a male are 2.82 times the females' odds of sending chat messages to a female vs. a male.  Since this measure only depends on the mean arc strength per arc--blocks, it could be interpreted as evidence of heterophily in the network, since it would mean $\mu_{MF}/ \mu_{MM} = 2.82 \mu_{FF}/ \mu_{FM}$.  However, its associated confidence interval contains the point one, which does not allow us to reject the hypothesis $\mu_{MF}/ \mu_{MM} = \mu_{FF}/ \mu_{FM}$.  We also obtain the interaction ratio $\hat m_{MF}/\hat m_{FM}=2.69$, with an associated confidence interval that allows us to say that the expected interaction from male to female is significatively larger than the expected interaction from female to male.  When discounting the asymmetry of the relationships between females and males from this interaction ratio, we obtain a measure of $0.99$ with a really sharp associated confidence interval.  Under the LASSB, this measure can be interpreted as evidence that the asymmetries in the relationships among the mixed--gender dyads nearly explain the imbalance of the interaction ratio, whereas the gregariousness and the attractiveness of the blocks jointly nearly compensate each other.

\begin{figure}[t]
\centering
  \centerline{\includegraphics[trim = 1cm .1cm 0cm 1cm, width=.425\linewidth]{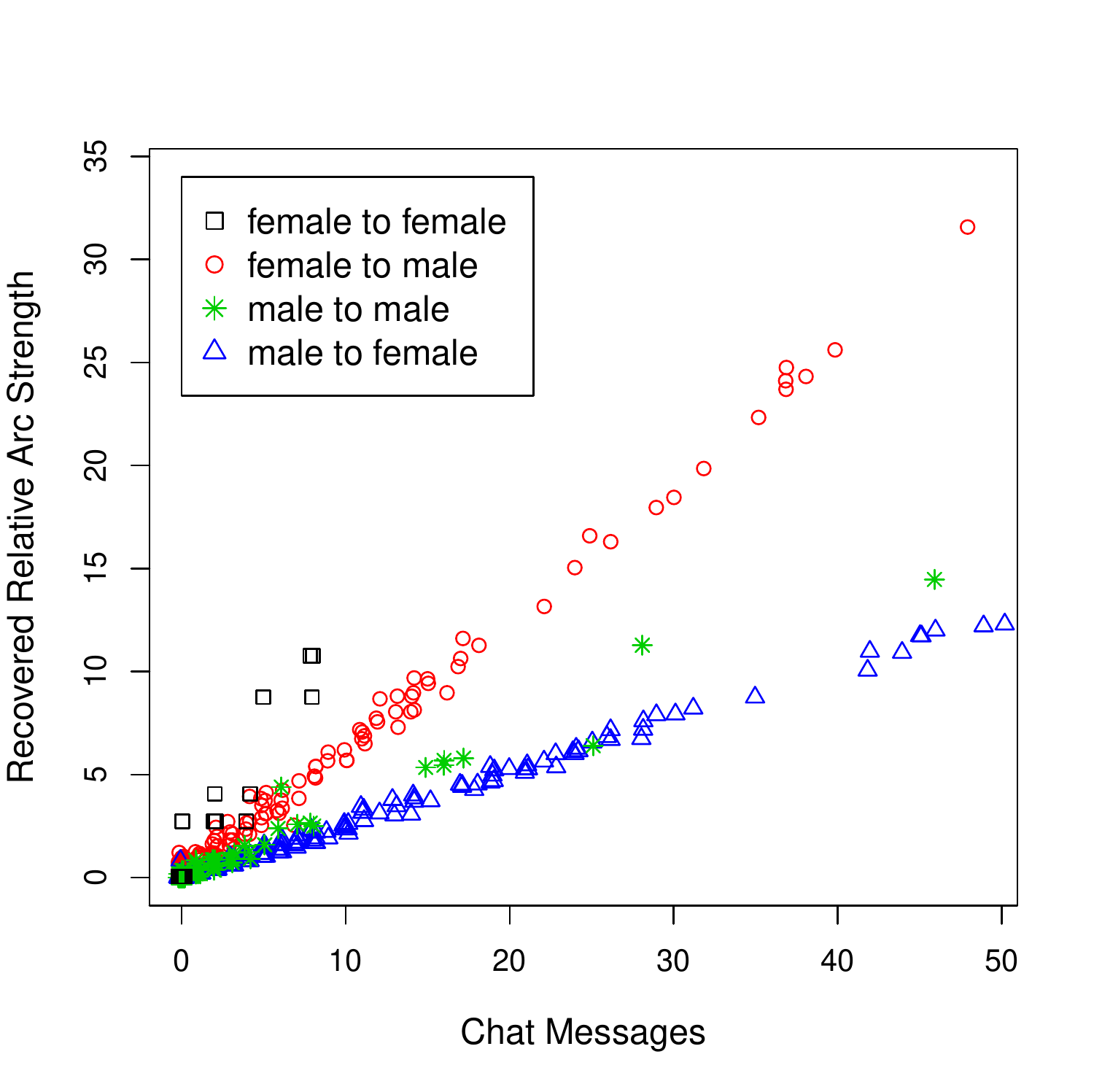}\hspace{.5cm}\includegraphics[trim = 0cm .1cm 1cm 2cm, width=.425\linewidth]{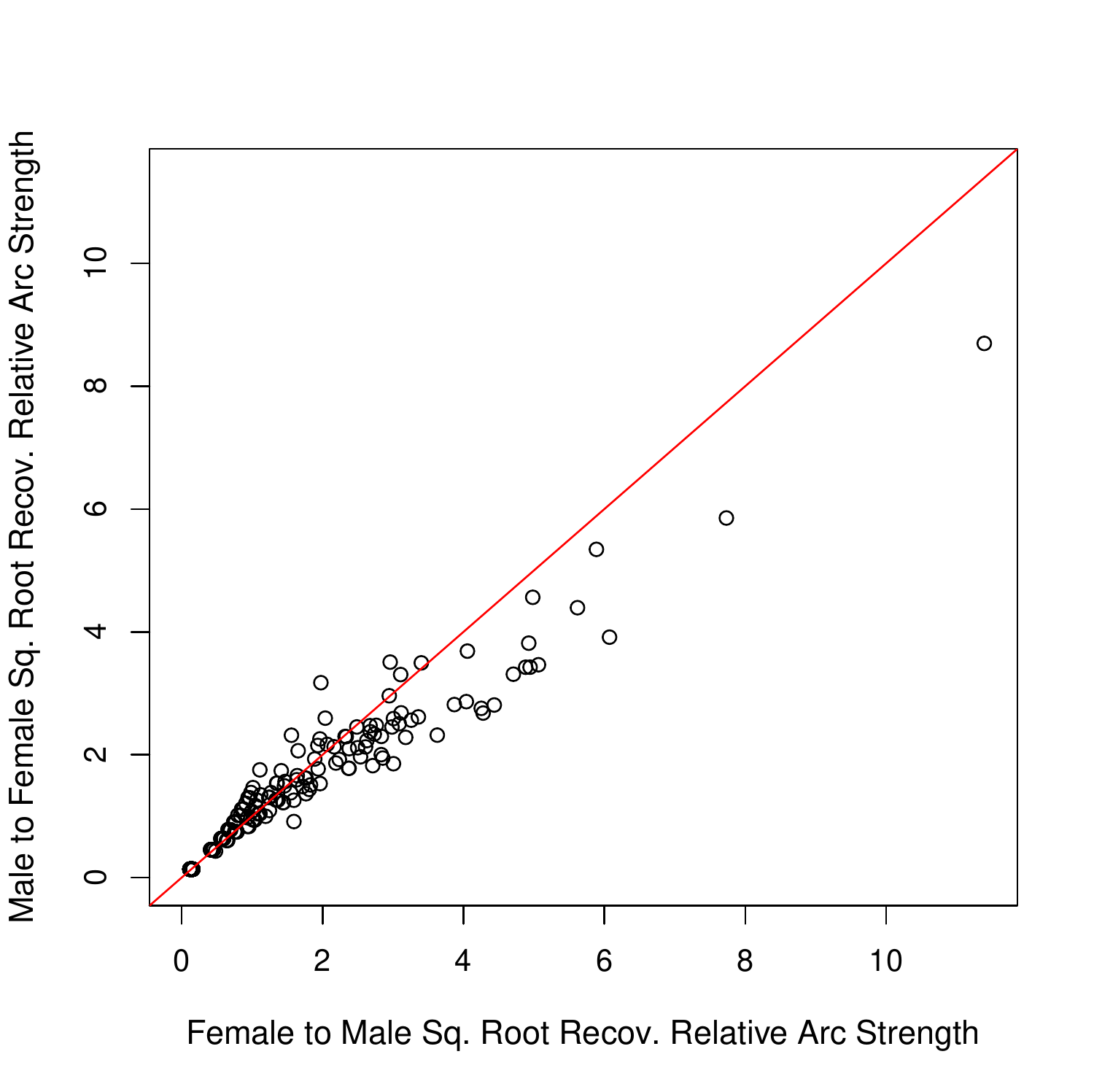}}
  \begin{minipage}[b]{0.85\textwidth}
  \caption{Left: number of chat messages per arc vs. recovered relative arc strength. Right: female to male vs. male to female recovered relative arc strengths, on the squared root scale.  The diagonal line represents equality between the two relative arc strengths coming from a dyad.
  }\label{f:rts}
\end{minipage}
\end{figure}

In the left hand side of Figure \ref{f:rts} we explore how the recovered relative arc strength changes compared to the actual amount of interaction, measured as number of chat messages.  We can see that as the number of chat messages increases, the relative arc strength increases at different rates for the four arc--blocks.  For instance, an interaction of 10 messages from female to female is relatively more important than 10 messages from a male to a female.  Finally, in the right hand side of Figure \ref{f:rts} we compare the two recovered relative arc strengths for mixed--gender dyads.  We can see that large values of female--to--male recovered relative arc strength are paired with lower values for the male--to--female counterpart, which contrasts with the observations that we made on chat messages exchanged between mixed--gender dyads (Figure \ref{f:sex_chats}).  Although males tend to send more chat messages to females than females to males, whenever a dyad has relatively strong arcs, the female--to--male arc tends to be relatively more important than the male--to--female one.

\subsection{Checking Goodness of Fit}\label{ss:GoF}

In order to check the goodness of fit of our model we follow the ideas of \cite{HunterGoodreauHandcock08} by comparing structural statistics of the observed network with the corresponding statistics on networks simulated from the fitted model.  We firstly check how our model fits to a dichotomized version of the network by studying the nodal distribution of \textit{binary outdegree}: $\sum_j I(X_{ij}>0)$, and \textit{binary indegree}: $\sum_j I(X_{ji}>0)$, where $I(\cdot)$ represents the indicator function.  We choose to study these statistics since in different applications, interaction counts are dichotomized assigning a link from node to node whenever there is some amount of interaction, and consequently it is reasonable to ask if our model predicts well these binary characteristics, despite it being designed for interaction networks (see \cite{ThomasBlitzstein11} for a discussion on consequences of dichotomizing valued networks).  We also study how the fitted LASSB predicts the distribution of \textit{valued outdegree}: $\sum_j X_{ij}$, and \textit{valued indegree}: $\sum_j X_{ji}$, since these two measures represent the total amount of interaction going \emph{from}, and \emph{to} certain user, respectively.  We also consider the distribution among dyads of \textit{absolute interaction difference}: $|X_{ij}-X_{ji}|$, since this measure reflects the dependencies of the arcs in a dyad.   Finally, since our model assumes dyadic independence, it is important to check whether this assumption is reasonable for the Kolkata friendships network.  A good way to check this assumption is by studying triads' characteristics, since this allows to detect transitivity effects in the network. Let the \textit{dyad interaction}  be $X_{i\wedge j} = X_{ij}+X_{ji}$.  The number of \textit{triangles at level $c$} is defined as $\sum_{i<j<k}I(X_{i\wedge j}>c)I(X_{i\wedge k}>c)I(X_{j\wedge k}>c)$, where $c$ represents a cutoff value.  We explore this measure for different cutoff values $c$.

\begin{figure}[!t]
    \centering
    \centerline{\includegraphics[trim = 0cm 0cm 0cm 0cm, width=.85\linewidth]{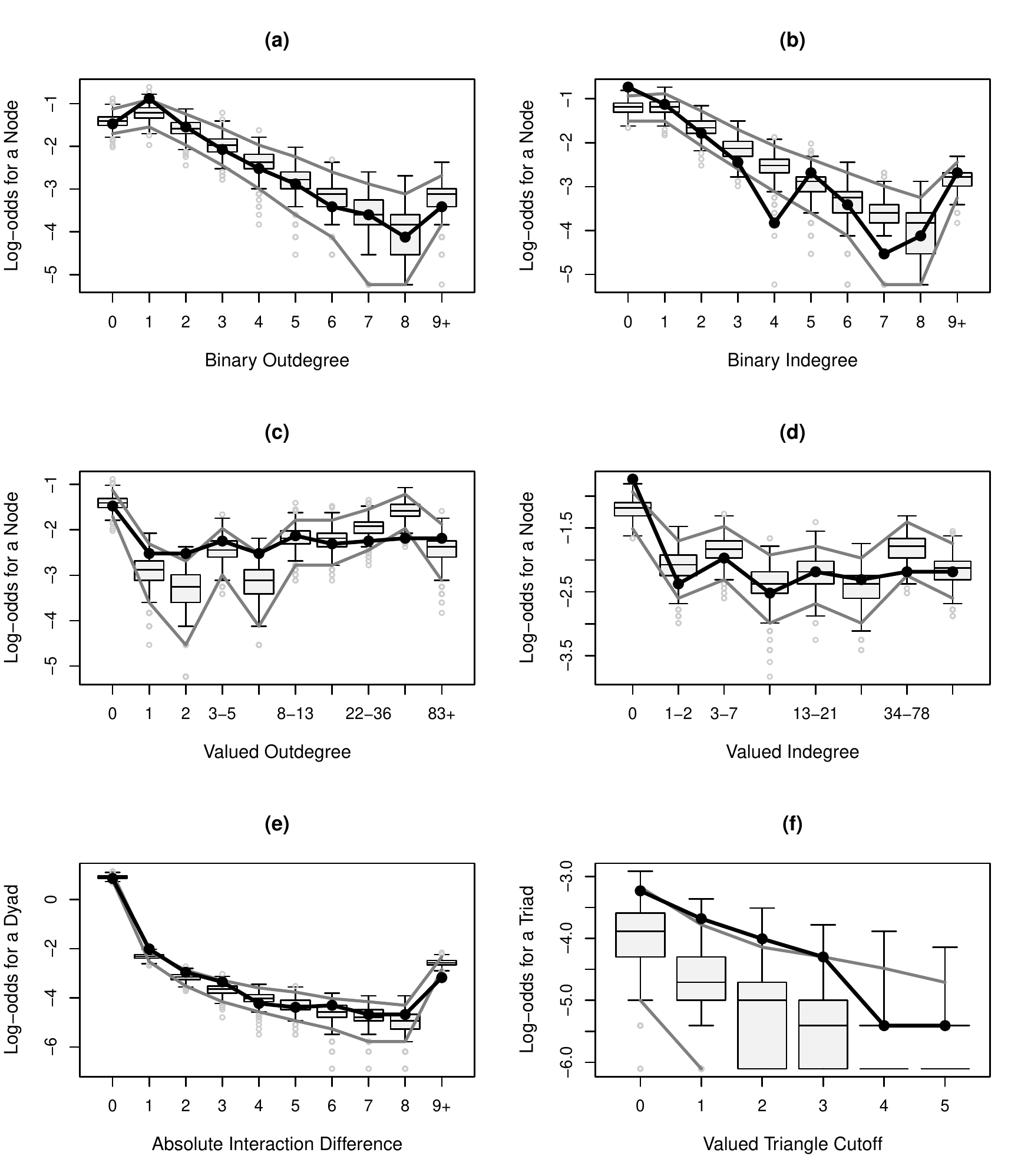}}
    \begin{minipage}[b]{0.85\textwidth}
    \caption{Summary of 1,000 simulations for checking the goodness of fit of the LASSB to the Kolkata interaction data.  In all plots the frequencies are presented in the log--odds scale, the black lines represent the observed frequencies in the Kolkata data, and the gray lines represent intervals containing 95\% of the simulated frequencies.  The boxplots represent the distribution of the frequencies obtained from the simulated networks.
    }
    \label{f:check1}
    \end{minipage}
\end{figure}

We generated 1,000 interaction networks from the fitted LASSB model, and we present the summarized results in Figure \ref{f:check1}, following the format of \cite{HunterGoodreauHandcock08}.  For each simulated network, and for the first five statistics mentioned above, we computed the proportion of nodes or dyads with their corresponding statistic being equal to a specific value or range of values, as specified in the horizontal axis of panels (a) to (e) in Figure \ref{f:check1}.  We also computed the proportion of triangles among friendship triads varying the cutoff $c$ from zero to six.  For the statistics on nodes, and on dyads, the distributions of their corresponding simulated proportions are explored using one box--plot on the log--odds scale for each value or range of values of the different statistics.  In the case of the proportion of triangles at different levels, one box--plot is presented for each cutoff $c$.  In Figure \ref{f:check1} the bold black lines represent the proportions observed in the Kolkata friendships network, and the gray lines represent intervals containing 95\% of the simulated proportions.  We say that we obtain a good fit of the model to certain network characteristic if the observed proportions fall within the range of variation obtained from the simulation.

We can see from panels (a) and (b) in Figure \ref{f:check1} that the LASSB fits well to the distribution of binary outdegree and binary indegree of the network, although the observed proportion of nodes having a binary indegree equal to four is low compared to the values obtained in the simulations.  Panel (c) of Figure \ref{f:check1} shows how the model fits to the distribution of valued outdegree.  We can see that the model tends to produce networks where the distribution of valued outdegree has a right tail heavier than in the observed network, although the observed values fall within the variation range predicted from the model.  Panel (d) in Figure \ref{f:check1} indicates that the fit of the model to the distribution of valued indegree is pretty good, although the proportion of nodes with observed zero valued indegree is a little high compared to the normal range of variation obtained from the model.  From panel (e) in Figure \ref{f:check1} we can observe that the model fits properly to the distribution of absolute interaction difference among friendship dyads, although the simulated proportions of dyads with an absolute interaction difference of nine or more tend to be larger than what we observe in the Kolkata network.  Finally, in panel (f) of Figure \ref{f:check1} we can see how the LASSB fits to the proportion of triangles at different cutoff levels.  We can see that the proportions of triangles in the Kolkata friendships network are in general larger than what is typically expected from the model.  This result was expected since our model does not capture transitivity effects.  Nevertheless, the fact that the observed proportions of triangles fall within the simulated variation range indicates that the transitivity effects in the network are not too large.

\section{Discussion}\label{s:discussion}

Our approach to arc/edge strength requires the construction of a null model that can be regarded as a scenario where there are no differential values of arc/edge strength.  We argued that this null model should include a notion of independence for the expected amounts of interaction, which in the case of directed interaction networks indicates that only the nodes' gregariousness and attractiveness explain the expected interaction counts.   The null model for the interaction network requires the complete specification of a distributional form.  In this article we explained why an independent Poisson process for each interaction count could be a sensible choice, although we expect other choices to be reasonable as well.  Arc/edge strength was therefore casted as a multiplicative departure from the expected amount of interaction under our null model.

We showed that taking each arc/edge strength as a fixed effect leads to models containing too many parameters, which is not useful for statistical analysis.  Using our ideal approach to arc/edge strength as a guideline, we mentioned a range of alternatives that aim to build parsimonious models by treating parameters as functions of covariates or as latent variables.  In particular, we developed a latent arc strength stochastic blockmodel (LASSB) for directed interaction networks, which takes arc strength as a latent variable, and jointly models the two arc strengths coming from a dyad, capturing dependencies in the interaction counts.  The LASSB further assumes the existence of blocks of nodes which are homogeneous with respect to their gregariousness and attractiveness, leading to a parsimonious way to explore the structure of the network.  Given that the LASSB does not involve node--specific fixed effects, it can be used to explore arc strength for a subset of dyads that may be the focus of interest, such as dyads with declared friendships in online social networks.

We saw how the proposed ideas allow to quantify asymmetries of the relationships between genders in the myGamma Kolkata network.  This approach also helped us to explore distributional characteristics of arc strength in the network, which is indicative of the strength of the online relationships between users of myGamma.  Although we saw that our model fitted properly to some characteristics of the observed network, it is clear that models relaxing the dyad independence assumption would capture more information about the structure of the network, such as the presence of transitivity in the interaction between users.  This article can be considered as a first step in the construction of appropriate methodologies that incorporate the notion of arc/edge strength in the modeling of interaction networks.

\section*{Acknowledgements}

The author thanks David S. Choi, Stephen E. Fienberg, Mike Finegold, Cosma Shalizi, and Andrew C. Thomas for helpful suggestions that contributed to many improvements of this work.  This research is supported by the Singapore National Research Foundation under its  International Research Centre @ Singapore Funding Initiative and administered by the IDM Programme Office.

\appendix

\section{Appendix: EM Algorithm for the LASSB}\label{a:EM}
We present an EM algorithm \citep{DempsterLairdRubin77} for fitting the LASSB via maximum likelihood.
\subsection{Complete Likelihood}

In the formulation of the LASSB in equation \eqref{eq:LASSB}, the gamma, and beta parts of the model are parameterized in terms of their means $\mu_{r\wedge s}$, and $\pi_{rs}$, such that their density functions are given by
\begin{equation*}
h_\Gamma(\lambda_{i \wedge j}) = \frac{1}{\Gamma(\nu_{r\wedge s})} \left(\frac{\nu_{r\wedge s}}{\mu_{r\wedge s}}\right)^{\nu_{r\wedge s}} \lambda_{i \wedge j}^{\nu_{r\wedge s}-1} \exp(-\nu_{r\wedge s} \lambda_{i \wedge j} / \mu_{r\wedge s} ),
\end{equation*}
and
\begin{equation*}
h_{\textrm{B}}(\rho_{ij}) = \frac{1}{\textrm{B}\big(\phi_{r\wedge s}\pi_{rs},\phi_{r\wedge s}(1-\pi_{rs})\big)} \rho_{ij}^{\phi_{r\wedge s}\pi_{rs}-1} (1-\rho_{ij})^{\phi_{r\wedge s}(1-\pi_{rs})-1}.
\end{equation*}
The density for $\lambda_{i \wedge j}$ is presented just for clarification, since we actually use the LASSB as in equation \eqref{eq:LASSB1}, where the gamma part has its mean fixed as one.  Now, let us define $\tau_{ij} = \theta\alpha_{r}\beta_{s}\mu_{r\wedge s}$ if $(i,j)\in B_{rs}$ in equation \eqref{eq:LASSB1}.  The complete likelihood of the LASSB is thus given by
\begin{eqnarray*}
\prod_{r\leq s}\prod_{\overset{i< j}{(i,j)\in B_{rs}}}&& \left\{ \frac{\nu_{r\wedge s}^{\nu_{r\wedge s}}}{\Gamma(\nu_{r\wedge s})\textrm{B}\left(\phi_{r\wedge s}\pi_{rs},\phi_{r\wedge s}(1-\pi_{rs})\right)}\frac{ \tau_{ij}^{x_{ij}} \tau_{ji}^{x_{ji}} }{ x_{ij}!x_{ji}!}  \rho_{ij} ^{x_{ij} +\phi_{r\wedge s}\pi_{rs}-1}(1-\rho_{ij})^{x_{ji} +\phi_{r\wedge s}(1-\pi_{rs})-1} \right. \\
&& \times \left. \exp\Big[-\gamma_{i \wedge j}\Big(\rho_{ij}(\tau_{ij}-\tau_{ji})+\tau_{ji}+\nu_{r\wedge s}\Big)\Big] \gamma_{i \wedge j} ^{x_{i\wedge j} + \nu_{r\wedge s}-1} \right\},
\end{eqnarray*}
where $x_{i\wedge j} = x_{ij}+x_{ji}$.  Note that the kernel of the conditional density of $\gamma_{i \wedge j}, \rho_{ij} | X=x$ can be obtained specifying the following hierarchical structure:
\begin{equation}\label{eq:postbeta}
\rho_{ij}|X=x,(i,j)\in B_{rs}  \sim \text{Beta}\left( \frac{x_{ij} +\phi_{r\wedge s}\pi_{rs}}{x_{i\wedge j} +\phi_{r\wedge s} }, x_{i\wedge j} +\phi_{r\wedge s}\right),
\end{equation}
\begin{equation}\label{eq:postgamma}
\gamma_{i \wedge j}|X=x, \rho_{ij}, (i,j)\in B_{rs} \sim \text{Gamma}\left( \frac{x_{i\wedge j} +\nu_{r\wedge s}}{\rho_{ij}(\tau_{ij}-\tau_{ji})+\tau_{ji}+\nu_{r\wedge s} }, x_{i\wedge j} +\nu_{r\wedge s}\right).
\end{equation}
This property is useful for deriving the expectation step of the EM algorithm.

In order to estimate the vector of parameters $\Phi$ involved in the model, we only need to take into account the part of the complete log--likelihood that involves $\Phi$, i.e.,
\begin{eqnarray}\label{eq:loglik}
l(\Phi; \rho, \gamma, x)&=& \sum_{r\leq s} \Biggr\{ b_{r\wedge s}\Big[\nu_{r\wedge s}\log \nu_{r\wedge s}-\log\Gamma(\nu_{r\wedge s})-\log\textrm{B}\big(\phi_{r\wedge s}\pi_{rs},\phi_{r\wedge s}(1-\pi_{rs})\big)\Big]\\
&& + \phi_{r\wedge s}\sum_{\overset{i< j}{(i,j)\in B_{rs}}} \Big(\pi_{rs}\log \rho_{ij}+(1-\pi_{rs})\log (1-\rho_{ij})\Big) - \nu_{r\wedge s}\sum_{\overset{i< j}{(i,j)\in B_{rs}}}(\gamma_{i\wedge j}-\log\gamma_{i\wedge j}) \nonumber\\
&& + \sum_{\overset{i\neq j}{(i,j)\in B_{rs}}}( x_{ij}\log \tau_{ij} - \gamma_{ij}\tau_{ij}) \Biggl\},\nonumber
\end{eqnarray}
where $b_{r\wedge s}$ denotes the number of dyads in $B_{r\wedge s}$.

\subsection{Expectation Step}
Using a vector of estimates $\Phi^{(t)}$ from iteration $t$, the EM algorithm requires for the E step the computation of
\begin{equation}\label{eq:E}
E_{\Phi^{(t)}} \left[l(\Phi; \rho, \gamma, X)|X=x\right] = \sum_{i<j} E_{\Phi^{(t)}} \left[l(\Phi; \rho_{ij}, \gamma_{i\wedge j}, X_{ij}, X_{ji})|X_{ij}=x_{ij}, X_{ji}=x_{ji}\right].
\end{equation}
However, by linearity of the expectation, we just need to compute the five expectations presented below.  Using \eqref{eq:postbeta} we find
\[\varrho_{ij}^{(t+1)} := E_{\Phi^{(t)}}(\log\rho_{ij}|X=x)=\psi(x_{ij} +\phi_{i\wedge j}^{(t)}\pi_{ij}^{(t)})-\psi(x_{i\wedge j} +\phi_{i\wedge j}^{(t)}),\]
\[\varrho_{ji}^{(t+1)} := E_{\Phi^{(t)}}(\log(1-\rho_{ij})|X=x)=\psi\big(x_{ji} +\phi_{i\wedge j}^{(t)}(1-\pi_{ij}^{(t)})\big)-\psi(x_{i\wedge j} +\phi_{i\wedge j}^{(t)}),\]
where $\psi(\cdot)$ represents the digamma function, and $\phi_{i\wedge j}=\phi_{r\wedge s}$, $\pi_{ij}=\pi_{rs}$ if $(i,j)\in B_{rs}$.  Using \eqref{eq:postbeta} and \eqref{eq:postgamma}, the law of total expectation, and integral representations of the hypergeometric and generalized hypergeometric functions, we obtain
\begin{eqnarray}\label{eq:Egamma}
\gamma_{i\wedge j}^{(t+1)}&:= &E_{\Phi^{(t)}}(\gamma_{i\wedge j}|X=x)\nonumber\\
&=&(x_{i\wedge j} +\nu_{i\wedge j}^{(t)})E_{\Phi^{(t)}}\Bigg(\frac{1}{\rho_{ij}\big(\tau_{ij}^{(t)}-\tau_{ji}^{(t)}\big)+\tau_{ji}^{(t)} +\nu_{i\wedge j}^{(t)}}\  \Big| X=x\Bigg)\nonumber\\
&=&\frac{x_{i\wedge j} +\nu_{i\wedge j}^{(t)}}{\tau_{ji}^{(t)} +\nu_{i\wedge j}^{(t)}} \,_2F_1\left(
\begin{array}{cc}
     1, & x_{ij} +\phi_{i\wedge j}^{(t)}\pi_{ij}^{(t)} \\
        & x_{i\wedge j} +\phi_{i\wedge j}^{(t)} \\
  \end{array};
\frac{\tau_{ji}^{(t)}-\tau_{ij}^{(t)}}{\tau_{ji}^{(t)}+\nu_{i\wedge j}^{(t)}}
\right),
\end{eqnarray}
and
\begin{eqnarray}\label{eq:Erhogamma}
\gamma_{ij}^{(t+1)}&:= &E_{\Phi^{(t)}}(\rho_{ij}\gamma_{i\wedge j}|X=x)\nonumber\\
&=&(x_{i\wedge j} +\nu_{i\wedge j}^{(t)})E_{\Phi^{(t)}}\Bigg(\frac{\rho_{ij}}{\rho_{ij}\big(\tau_{ij}^{(t)}-\tau_{ji}^{(t)}\big)+\tau_{ji}^{(t)} +\nu_{i\wedge j}^{(t)}}\  \Big| X=x\Bigg)\nonumber\\
&=& \frac{x_{i\wedge j} +\nu_{i\wedge j}^{(t)}}{\tau_{ij}^{(t)}-\tau_{ji}^{(t)}}\left[1-\,_2F_1\left(
\begin{array}{cc}
     1, & x_{ij} +\phi_{i\wedge j}^{(t)}\pi_{ij}^{(t)} \\
        & x_{i\wedge j} +\phi_{i\wedge j}^{(t)} \\
  \end{array};
\frac{\tau_{ji}^{(t)}-\tau_{ij}^{(t)}}{\tau_{ji}^{(t)}+\nu_{i\wedge j}^{(t)}}
\right)\right],
\end{eqnarray}
where $\,_2F_1(\cdot)$ represents the hypergeometric function \cite[see][]{Daalhuis10}, and $\nu_{i\wedge j}=\nu_{r\wedge s}$ if $(i,j)\in B_{rs}$.  Similarly we obtain
\begin{eqnarray}
\varsigma_{ij}^{(t+1)} &:= & E_{\Phi^{(t)}}(\log\gamma_{i\wedge j}|X=x) \nonumber \\
&=& \psi(x_{i\wedge j} +\nu_{i\wedge j}^{(t)})- E_{\Phi^{(t)}}\Big(\log\Big[\rho_{ij}\big(\tau_{ij}^{(t)}-\tau_{ji}^{(t)}\big)+\tau_{ji}^{(t)} +\nu_{i\wedge j}^{(t)}\Big]\  \Big| X=x \Big) \label{eq:varsigmaMC}\\
&\overset{!}{=}&\psi\big(x_{i\wedge j} +\nu_{i\wedge j}^{(t)}\big)-\log\big(\tau_{ji}^{(t)}+\nu_{i\wedge j}^{(t)}\big)\label{eq:varsigmahypergeo}\\
&&+\frac{\big(x_{ij} + \phi_{i\wedge j}^{(t)}\pi_{ij}^{(t)}\big)\big(\tau_{ji}^{(t)}-\tau_{ij}^{(t)}\big)}{\big(x_{ij} +x_{ji} +\phi_{i\wedge j}^{(t)}\big)\big(\tau_{ji}^{(t)}+\nu_{i\wedge j}^{(t)}\big)} \,_3F_2\left(
\begin{array}{ccc}
    1, & 1, & x_{ij} +\phi_{i\wedge j}^{(t)}\pi_{ij}^{(t)}+1 \\
     & 2, & x_{i\wedge j} +\phi_{i\wedge j}^{(t)}+1 \\
  \end{array};
\frac{\tau_{ji}^{(t)}-\tau_{ij}^{(t)}}{\tau_{ji}^{(t)}+\nu_{i\wedge j}^{(t)}}
\right),\nonumber
\end{eqnarray}
where $\,_3F_2(\cdot)$ represents the generalized hypergeometric function \cite[see][]{AskeyDaalhuis10}.  Equation \eqref{eq:varsigmahypergeo} holds only if $|(\tau_{ji}^{(t)}-\tau_{ij}^{(t)})/(\tau_{ji}^{(t)}+\nu_{i\wedge j}^{(t)})|<1$, otherwise we compute \eqref{eq:varsigmaMC} using a Monte Carlo approximation for the unevaluated expectation, taking a large random sample from a beta distribution with parameters as in equation \eqref{eq:postbeta}, but using $\phi_{r\wedge s}^{(t)}$, and $\pi_{rs}^{(t)}$.  Implementations of the hypergeometric and generalized hypergeometric functions are available in the R package \verb"hypergeo" \citep{hypergeo}.

\subsection{Maximization Step}\label{ss:Msection}

For the M step we need to find
\begin{equation*}
\Phi^{(t+1)} = \underset{\Phi}{\arg\max}\Big\{E_{\Phi^{(t)}} \left[l(\Phi; \rho, \gamma, X)|X=x\right]\Big\}.
\end{equation*}
From equation \eqref{eq:loglik} we can see that the maximization over $\Phi$ can be obtained independently over three subsets of parameters: $\{\nu_{r\wedge s}; r,s=1,\dots,S\}$, $\{\pi_{rs},\phi_{r\wedge s}; r,s=1,\dots,S\}$, and $\{\theta,\alpha_r,\beta_s,\mu_{r\wedge s}; r,s=1,\dots,S\}$.

In order to maximize with respect to $\{\theta,\alpha_r,\beta_s,\mu_{r\wedge s}; r,s=1,\dots,S\}$, note that these parameters are only involved in the Poisson part of the complete likelihood, which allows to estimate them from a Poisson log--linear model of quasi--symmetry with offset, i.e., $X_{ij}\overset{ind}{\sim} \text{Poisson}(m_{ij})$, where
\begin{equation}
\log m_{ij} = \eta + \eta_r^\mathcal{I} + \eta_s^\mathcal{J} + \eta_{rs}^\mathcal{IJ} + \log\gamma_{ij}^{(t+1)},
\end{equation}
if $(i,j)\in B_{rs}$, with the quasi--symmetry constraint $\eta_{rs}^\mathcal{IJ}=\eta_{sr}^\mathcal{IJ}$, and the usual $\eta_1^\mathcal{I} = \eta_1^\mathcal{J} = 0$, and $\eta_{r1}^\mathcal{IJ} = \eta_{1s}^\mathcal{IJ} = 0$ for all $r,s$.  This formulation allows to take advantage of software built--in procedures to estimate generalized linear models via maximum likelihood (e.g., \verb"glm" in R).  We use the functional invariance property of MLEs to obtain $\{\theta^{(t+1)},\alpha_r^{(t+1)},\beta_s^{(t+1)},\mu_{r\wedge s}^{(t+1)}; r,s=1,\dots,S\}$ by exponentiating the corresponding coefficients of the log--linear model, e.g., $\theta^{(t+1)} = \exp(\eta^{(t+1)})$ and so on.  We take $\tau_{ij}^{(t+1)}:=\theta^{(t+1)}\alpha_r^{(t+1)}\beta_s^{(t+1)}\mu_{r\wedge s}^{(t+1)}$ if $(i,j)\in B_{rs}$.

In order to maximize over $(\pi_{rs},\phi_{r\wedge s})$ we need to maximize the function
\begin{eqnarray*}
f^{(t+1)}(\pi_{rs},\phi_{r\wedge s})&=&- b_{r\wedge s}\log\textrm{B}\big(\phi_{r\wedge s}\pi_{rs},\phi_{r\wedge s}(1-\pi_{rs})\big) + \phi_{r\wedge s}\sum_{\overset{i< j}{(i,j)\in B_{rs}}} \Big(\pi_{rs}\varrho_{ij}^{(t+1)}+(1-\pi_{rs})\varrho_{ji}^{(t+1)}\Big),
\end{eqnarray*}
and we take $(\pi_{rs}^{(t+1)},\phi_{r\wedge s}^{(t+1)}) = \underset{\pi_{rs},\phi_{r\wedge s}}{\arg\max}\ f^{(t+1)}(\pi_{rs},\phi_{r\wedge s})$.  However, when $r=s$, we fix $\pi_{rs}=0.5$.  Finally, the objective function to maximize over $\nu_{r\wedge s}$ reduces to
\begin{eqnarray*}
g^{(t+1)}(\nu_{r\wedge s})&=&b_{r\wedge s}\Big[\nu_{r\wedge s}\log \nu_{r\wedge s}-\log\Gamma(\nu_{r\wedge s})\Big]- \nu_{r\wedge s}\sum_{\overset{i< j}{(i,j)\in B_{rs}}}(\gamma_{i\wedge j}^{(t+1)}-\varsigma_{ij}^{(t+1)}),
\end{eqnarray*}
and we find $\nu_{r\wedge s}^{(t+1)} = \underset{\nu_{r\wedge s}}{\arg\max}\ g^{(t+1)}(\nu_{r\wedge s})$.  All the previous maximizations can be obtained using an iterative method such as the Nelder--Mead algorithm.

\subsection{Starting Values}

We need to provide the EM algorithm with starting values $\Phi^{(0)}$.  We propose to compute $\Phi^{(0)}$ using some reasonable initial measures  $\gamma^{(0)}_{i\wedge j}$, and $\rho^{(0)}_{ij}$.  Let $x^{*}_{ij}=x_{ij}+\varepsilon$ ($\varepsilon$ is, e.g., 0.05), $x^{*}_{i\wedge j}=x^{*}_{ij}+x^{*}_{ji}$, and $\bar x^{*}_{r\wedge s}=(1/b_{r\wedge s})\sum_{(i,j)\in B_{rs}}x^{*}_{i\wedge j}$, for $r\leq s$.  We add a small $\varepsilon$ to $x_{ij}$ in order to avoid initial zero $\gamma^{(0)}_{ij}$, since our proposal for computing $\Phi^{(0)}$ does not work otherwise.  We take $\gamma^{(0)}_{i\wedge j} = x^{*}_{i\wedge j}/\bar x^{*}_{r\wedge s}$, for $(i,j)\in B_{rs}, r\leq s$, since this measure captures the total interaction of the dyad, and it has mean one for each dyad--block.  We also take $\rho^{(0)}_{ij} = x^{*}_{ij}/x^{*}_{i\wedge j}$ since this captures the share of the arc $(i,j)$ in the dyad interaction.  Finally, $\gamma^{(0)}_{ij}=\gamma^{(0)}_{i\wedge j}\rho^{(0)}_{ij}=x^{*}_{ij}/\bar x^{*}_{r\wedge s}$.

In order to find $\nu_{r\wedge s}^{(0)}$, $\pi_{rs}^{(0)}$, and $\phi_{r\wedge s}^{(0)}$ we propose to use a method of moments approach.  From the parametrization presented in equation \eqref{eq:LASSB1} it is easy to check that $Var(\gamma_{i \wedge j})=1/\nu_{r\wedge s}$, for $(i,j)\in B_{rs}$, from which we take
\[\nu_{r\wedge s}^{(0)}=\left[\frac{1}{b_{r\wedge s}}\sum_{\overset{i< j}{(i,j)\in B_{rs}}} \left(\gamma^{(0)}_{i \wedge j}-1\right)^2\right]^{-1}.\]

For $r<s$, we take $\pi_{rs}^{(0)}=(1/b_{r\wedge s})\sum_{(i,j)\in B_{rs}}\rho^{(0)}_{ij}$ (remember $\pi_{ss}=0.5$ is fixed).  It is also easy to check $E[\rho_{ij}(1-\rho_{ij})]=\pi_{rs}(1-\pi_{rs})\phi_{r\wedge s}/(1+\phi_{r\wedge s})$, for $(i,j)\in B_{rs}$, and hence
\[\phi_{r\wedge s}=\frac{E\big[\rho_{ij}(1-\rho_{ij})\big]}{\pi_{rs}(1-\pi_{rs})-E\big[\rho_{ij}(1-\rho_{ij})\big]},\]
from which we take
\[\phi_{r\wedge s}^{(0)}=\frac{(1/b_{r\wedge s})\sum_{i<j, (i,j)\in B_{rs}}\rho^{(0)}_{ij}(1-\rho^{(0)}_{ij})}{\pi^{(0)}_{rs}(1-\pi^{(0)}_{rs})-(1/b_{r\wedge s})\sum_{i<j,(i,j)\in B_{rs}}\rho^{(0)}_{ij}(1-\rho^{(0)}_{ij})}.\]
Note that using this approach to find $\phi_{s\wedge s}^{(0)}$ is appropriate, since it does not actually require an ordering among $i$ and $j$.  Finally, we take $\{\theta^{(0)}, \alpha^{(0)}_{r}, \beta^{(0)}_{s}, \mu^{(0)}_{r\wedge s}; r,s=1,\dots,S\}$ from a Poisson log--linear model of quasi--symmetry as in Section \ref{ss:Msection}, taking $\log \gamma^{(0)}_{ij}$ as an offset.

\bibliographystyle{APA}
\bibliography{biblionet}

\end{document}